\documentclass[10pt,twocolumn,letterpaper]{article}

\usepackage{cvpr}              
\usepackage{hyperref}
\usepackage{url}
\usepackage[abs]{overpic} 
\usepackage{booktabs} 
\usepackage{bm}
\usepackage{algorithm}
\usepackage[table]{xcolor}
\usepackage[noend]{algpseudocode} 
\usepackage{adjustbox}
\usepackage{multirow}
\usepackage{float}
\usepackage[absolute,overlay]{textpos}
\usepackage[accsupp]{axessibility}

\algrenewcommand\algorithmiccomment[1]{\hfill\(\triangleright\)\,#1}

\newcommand{\A}{\mathcal{A}}
\newcommand{\pre}{\textup{pre}}
\newcommand{\out}{\textup{out}}

\let \bs = \boldsymbol

\definecolor{cvprblue}{rgb}{0.21,0.49,0.74}

\def\paperID{}
\def\confName{CVPR}
\def\confYear{2026}

\newcommand{\best}{\cellcolor{red!25}} 
\newcommand{\second}{\cellcolor{orange!25}}

\newcommand{\bestbox}{\raisebox{0.1ex}{\colorbox{red!25}{\phantom{\rule{0.4em}{0.4em}}}}}
\newcommand{\secondbox}{\raisebox{0.1ex}{\colorbox{orange!25}{\phantom{\rule{0.4em}{0.4em}}}}}

\title{SparseOIT: Improving Order-Independent Transparency 3DGS \\ via Active Set Method}

\author{
Wentao Yang$^{1,2}$ \quad
Fanzhen Kong$^{2}$ \quad
Zejian Kang$^{1,2}$ \quad
Xiangru Huang$^{2\dag}$\\
$^{1}$Zhejiang University \quad
$^{2}$Westlake University \\
}

\begin{document}

\twocolumn[{%
\renewcommand\twocolumn[1][]{#1}%
\maketitle
\includegraphics[width=\linewidth]{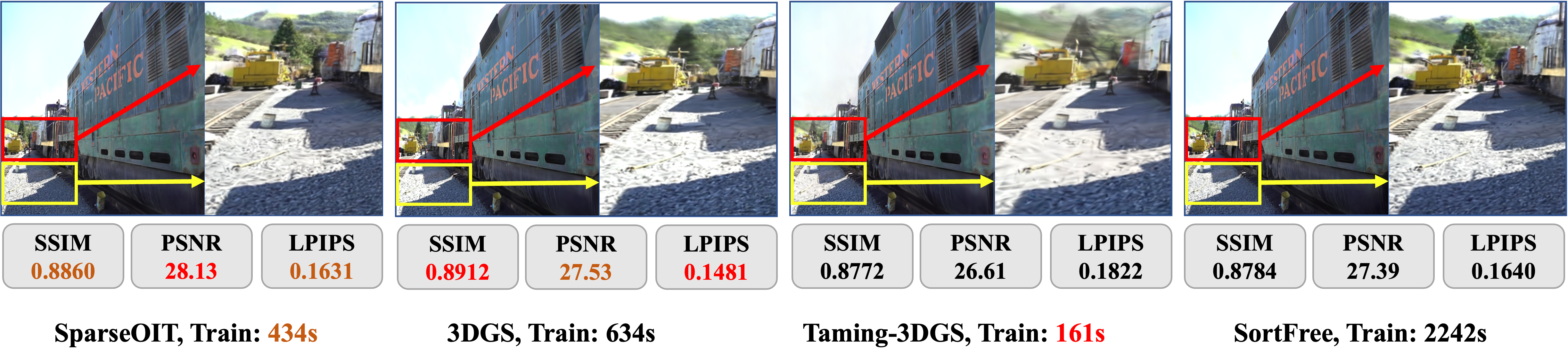}
\captionof{figure}{We introduce the SparseOIT, an improved Order-Independent Transparency based 3DGS~\cite{kerbl3Dgaussians} reconstruction method that achieves significant improvement over the existing OIT-based methods such as SortFree~\cite{hou2024sort} and comparable performance of state-of-the-art volumetric rendering based methods such as 3DGS~\cite{kerbl3Dgaussians} and Taming-3DGS~\cite{taminggs}. The rendering quality metrics shown in this figure are computed for a single view.
}
\vspace{0.8cm}
\label{fig:teaser}

}]

\begingroup
\renewcommand{\thefootnote}{\textdagger} 
\footnotetext{\hangindent=1.8em \hangafter=1 Corresponding author (huangxiangru@westlake.edu.cn).}
\endgroup

\thispagestyle{plain}
\begin{abstract}
3D Gaussian Splatting (3DGS) has received tremendous popularity over the past few years due to its photorealistic visual appearance. However, 3DGS uses volumetric rendering that is not suitable for objects with non-lambertian or transparent materials. To remedy this issue, a family of Order-Independent Transparency (OIT) rendering methods propose to remove or modify the depth sorting step in the 3DGS rendering equation. However, the potential of OIT-based method is still underexplored. In this paper, we observe that the OIT modifications to the rendering equation significantly reduce the inter-independence among individual gaussian splats, resulting in very sparse variable dependencies that can be harnessed by specific optimization techniques such as active set method. To this end, we propose \textbf{SparseOIT}, an OIT-based 3DGS reconstruction algorithm that maintains an active set of gaussian splats and enjoys an acceleration ratio that is proportional to the potential sparsity. SparseOIT is designed by jointly considering the OIT rendering equation, the reconstruction algorithm and the geometric regularization. Through extensive experiments, we demonstrate that SparseOIT outperforms existing methods in the OIT-family by a large margin and also achieves comparable performance to the state-of-the-art 3DGS reconstruction methods based on volumetric rendering. Project page: \url{https://wentaoyang19.github.io/SparseOIT.github.io/}
\end{abstract}    

\section{Introduction}
\label{sec:intro}

3D Gaussian Splatting (3DGS) has received an overwhelming popularity over the past few years due to its photorealistic visual appearance. Typically, 3DGS-based reconstruction algorithms use a set of gaussian splats and associated attributes such as spherical harmonics to represent the geometry and the appearance. Following Neural Radiance Field (NeRF), the gaussian splats are rendered according to the volumetric rendering equation. Informally, the key idea is to sequentially order gaussian splats according to their distance to the camera center and to simulate the effect where gaussian splats closer to the camera occludes the other splats. During optimization, as the gaussian splats are iteratively updated, this sorting step can lead to non-smooth gradient if the order of gaussian splats is changed. It also creates poping artifacts when the order of gaussian splats is changed in a novel view point. To remedy this issue, a family of Order-Independent Transparancy (OIT) rendering methods is proposed to remove the sorting step. For example, similar in the sort-free rendering paper, gaussian splats are weighted by their depth $d$ and the view direction $\bs{r}$, according to the following equation:
\begin{equation}
    w(d, \bs{r}) = \max \left( 0,1-\frac{d}{\sigma}\right)v(\bs{r})
\label{eq:sf_weight}
\end{equation}
where $\sigma$ is a learnable parameter of scene and $v(\bs{r}) \in \mathbb{R}^{4^2\times 3}$ is a degree-3 spherical harmonics (SH) that compensates for view-dependent appearance.

Even with the advancement in removing the sorting step, OIT-based methods is still limited from being widely used due to several challenges: 1) the computation time is significantly longer compared to 3DGS methods that use volumetric rendering; 2) in the experiments, the OIT-based methods suffer from much lower rendering quality in novel viewpoints. In this paper, we improve OIT-based methods by designing a reconstruction algorithm that jointly considers two ingredients: 1) the rendering equation, 2) the algorithm as well as its GPU implementation. 

Looking at the OIT rendering equation, when the sorting step is removed, gaussian splats are much more isolated, enabling optimization algorithms to individually optimize each gaussian splats. Similar to the 80/20 rule, an observation of the 3DGS optimization procedure is: when a scene or an object is being reconstructed, simple structures (such as tables, walls) with simple textures take up a large number of splats but only requires a small amount of iterations to converge; while small objects with complicated geometry is represented by a small number of splats but requires a large number of iterations. This observation suggests that during the optimization procedure, \emph{a large fraction of splats can be frozen for a large fraction of iterations}. The removing of the sorting step effectively allows us to convert this observation into actual benefits in terms of faster convergence or reduced running time. 

Specifically, we adopt the active set method and maintain a set of active gaussian splats. The optimization is performed only on the active set, which leads to an acceleration ratio that is proportional to the potential sparsity. To efficiently update the active set, we rely on subsampling methods to periodically estimate the gradients of gaussian splats with only sublinear complexity computations. The experiments suggest that the active set method significantly reduces the running time without significant affect on the novel-view rendering quality.

In addition, the 3DGS community has developed multiple GPU acceleration techniques such as Taming-3DGS~\cite{taminggs}. To reduce the performance gap between volumetric rendering based methods and OIT-based methods, we integrate the GPU acceleration techniques into the design of our algorithm, which significantly improves the performance.

To summarize, our contributions are
\begin{itemize}
    \item We propose to use active set method that is capable of fully harnessing the sparsity of gaussian activeness for OIT-based method during training, leading to strong acceleration in training time.
    \item We integrate popular GPU acceleration techniques into our algorithm design. Our algorithm achieves comparable performance to the state-of-the-art volumetric rendering methods and while significantly improves existing OIT-based methods.
\end{itemize}
\section{Related Works}
\label{sec:related_work}
The related works are reviewed from three perspectives. We first review volume rendering as the fundamental formulation underlying early 3DGS methods in Sec.~\ref{subsec:related_rendering}. We then introduce Order-Independent Transparency in Sec.~\ref{subsec:related_OIT} as an alternative formulation for transparency compositing without explicit sorting, and finally summarize representative acceleration techniques for improving the training and rendering efficiency of 3DGS in Sec.~\ref{subsec:related_acceleration}.

\subsection{Volume Rendering}
\label{subsec:related_rendering}

Early image-based rendering methods (e.g.,~\cite{debevec1998ibr}) and point-based systems such as QSplat~\cite{rusinkiewicz2000qsplat} demonstrated the effectiveness of explicit scene primitives for efficient rendering and novel view synthesis. The emergence of NeRF~\cite{mildenhall2020nerf} shifted the paradigm toward implicit volumetric rendering, enabling photorealistic novel-view synthesis through continuous radiance fields. Subsequent variants, including Mip-NeRF~\cite{barron2021mipnerf}, Mip-NeRF~360~\cite{barron2022mipnerf360}, and Instant-NGP~\cite{mueller2022instantngp}, improved anti-aliasing, scene scalability, and training speed, respectively. Explicit voxel-based radiance fields such as Plenoxels~\cite{yu2022plenoxels} and DVGO~\cite{sun2022dvgo} further reduced rendering cost by avoiding per-sample MLP inference. Despite their progress, NeRF-style methods often incur substantial computational or memory overhead. In contrast, 3DGS~\cite{kerbl3Dgaussians} adopts explicit anisotropic Gaussian primitives and a differentiable splatting rasterizer, achieving real-time rendering while maintaining high visual fidelity. The original 3DGS~\cite{kerbl3Dgaussians} relies on strict per-pixel depth sorting to ensure correct alpha compositing during rendering. However, explicit sorting incurs substantial computational overhead and may also lead to temporal artifacts such as flickering and popping~\cite{radl2024stopthepop}. These limitations motivate OIT as an alternative rendering formulation. 

\subsection{Order-Independent Transparency in 3D Gaussian Splatting}
\label{subsec:related_OIT}
Order-Independent Transparency (OIT) provides a well-established solution for rendering semi-transparent effects, such as smoke, clouds, and hair, without requiring strict depth sorting. To overcome the non-commutativity of the classical OVER operator~\cite{Thomas1984over}, OIT introduces several efficient approximations, including k-buffer~\cite{Bavoil2007KBuffer}, stochastic transparency~\cite{Enderton2010StochasticTransparency}.
Among these approaches, the weighted blended OIT method proposed by McGuire and Bavoil~\cite{mcguire2013weighted} is particularly influential, as it replaces the conventional non-commutative alpha blending formulation with a commutative weighted compositing scheme. Recently, OIT has also been applied to Gaussian-based representations. Sort-Free~\cite{hou2024sort} proposes a weighted-sum blending formulation that approximates standard alpha compositing in a commutative manner, enabling sorting-free rendering. StochasticSplats~\cite{kheradmand2025stochastic} leverages stochastic sampling of Gaussian contributions to unbiasedly resolve transparency without sorting, while Hybrid Transparency~\cite{hahlbohm2025efficient} combines accuracy and efficiency by depth-sorting only the top-K high-contribution fragments and applying weighted blending to the rest. These OIT strategies can effectively reduce rendering overhead and improve temporal stability. However, their effectiveness in improving both training and rendering efficiency remains limited.

\subsection{Acceleration Techniques for 3D Gaussian Splatting}
\label{subsec:related_acceleration}

Recent researches have demonstrate that the 3DGS pipeline can be significantly accelerated. Some works focus on reducing the number of Gaussians. C3DGS~\cite{lee2024c3dgs} introduces a learnable mask to eliminate Gaussians that contribute minimally to the final rendering quality, Speedy-Splat~\cite{HansonSpeedy} applies both soft and hard pruning techniques for accelerating. Other works aim to minimize computational overhead and memory consumption through Gaussian pruning, quantization, and efficient training strategies. EAGLES~\cite{Sharath2024eagles} adopts quantized embeddings to reduce per-point memory usage and employs a coarse-to-fine training strategy. AdR-Gaussian~\cite{xzwang2024adrgaussian} accelerates 3DGS by culling of Gaussian-tile pairs with low opacity. Additionally, Taming-3DGS~\cite{taminggs} demonstrates that high-quality 3DGS can be achieved even under constrained computational budgets by carefully balancing Gaussian growth, resource scheduling, and rendering-time approximation, further highlighting the importance of efficient design choices in accelerating 3DGS pipelines. In terms of rendering, Balanced3DGS~\cite{gui2024balanced} addresses GPU workload imbalance via Gaussian-wise parallelism and fine-grained tiling. These advances collectively push 3DGS toward scalable deployment in real-time and resource-constrained environments. Some works focus on optimization method, like 3DGS-LM~\cite{hoellein_2024_3dgslm} and 3DGS2~\cite{lan20253dgs2}. 3DGS-LM~\cite{hoellein_2024_3dgslm} replaces gradient descent with second-order Levenberg-Marquardt updates to achieve faster convergence. 3DGS2~\cite{lan20253dgs2} proposes a near second-order training scheme for 3D Gaussian Splatting by formulating per-attribute local Newton updates instead of relying on global SGD. Another work focus on acceleration on image, FreGS~\cite{zhang2024fregs} introduces a progressive frequency regularization scheme for 3D Gaussian Splatting, enforcing consistency between rendered and ground-truth images in the Fourier domain and annealing the regularization from low to high frequencies in a coarse-to-fine manner. DashGaussian~\cite{chen2025dashgaussian} proposes a frequency-aware resolution schedule and synchronized primitive growth to converge in under 200 seconds. However, acceleration methods based on residual updates over 3D Gaussian primitives remain largely unexplored.

\section{Preliminary}
\label{sec:Pre}
In this section, we provide a brief review of 3D Gaussian Splatting~\cite{kerbl3Dgaussians} (Sec.~\ref{subsec:3dgs}) and the OIT-based methods (Sec.~\ref{subsec:OIT}).

\subsection{3D Gaussian Splatting}
\label{subsec:3dgs}
3D Gaussian Splatting (3DGS) utilizes a set of 3D Gaussians $\mathcal{G}=\{\bm{\mu}_{i}, \bm{q}_i, \bm{s}_i, o_{i}, \bm{h}_i| i = 1, \ldots, N \}$ to represent the scene with splatting technique for rendering. The geometry attributes of 3D Gaussian consists of 
center position $\bm{\mu} \in \mathbb{R}^{3}$, unit quaternion $\bm{q} \in \mathbb{R}^{4}$ (for rotation), and scale $\bm{s} \in \mathbb{R}^{3}$. Additionally, the appearance attributes consists of opacity $o_i \in \mathbb{R}^+$, and the RGB color of each view direction encoded with degree-3 spherical harmonics (SH) $\bm{h}_i \in \mathbb{R}^{4^2\times 3}$. 

The 3D Gaussian $\bm{g}_{i}$ is defined as:
\begin{equation}
\bm{g}_{i}(\bm{x}) = e^{-\frac{1}{2} (\bm{x}-\bm{\mu})^T \bm{\Sigma}^{-1} (\bm{x}-\bm{\mu})}, 
\label{eq:3d_gaussian}
\end{equation}
where $\bm{\Sigma} = \bm{R}\bm{S}\bm{S}^{T}\bm{R}^{T}$ is covariance matrix, which can be derived from the rotation matrix $\bm{R}$ (derived from $\bm{q}$) and scaling matrix $\bm{S}$ (derived from $\bm{s}$). 

The rendering equation in~\cite{kerbl3Dgaussians} is defined by:
\begin{equation}
    \bm{C} = \sum_{i} T_i \alpha_i \bm{c_i},~\quad T_i = \prod_{j=1}^{i-1} \left(1-\alpha_j\right),
\label{eq:3dgs_rendering}
\end{equation}
where $\bm{c_i} \in \mathbb{R}^3$ is the learned color, and $\alpha_{i}$ is the final opacity.
The $\bm{c_i}$ is a learnable parameter, which can be derived by:
\begin{equation}
\bm{c_i} = \mathrm{Y}\left(\left\|\bm{f}-\bm{\mu}\right\|,\bm{h}_i\right), 
\label{eq:3dgs_color}
\end{equation}
where $\bm{f}$ is focal point and $\mathrm{Y}$ is spherical harmonic function. 
The final opacity $\alpha_{i}$ is denoted as:
\begin{equation}
\alpha_{i} = o_{i} \cdot \left({-\frac{1}{2} \left(\bm{x}_{i}^{\prime}-\bm{\mu}_{i}^{\prime}\right)^T {\bm{\Sigma}_{i}^{\prime}}^{-1} \left(\bm{x}_{i}^{\prime}-\bm{\mu}_{i}^{\prime}\right)}\right), 
\label{eq:3dgs_alpha}
\end{equation}
where $\bm{x}_{i}^{\prime}$ and $\bm{\mu}_{i}^{\prime}$ are projected and center position in 2D space, and $o_{i}$ is the opacity. The 2D covariance matrix $\bm{\Sigma}_{i}^{\prime}$ can be derived by:
\begin{equation}
\bm{\Sigma}^{\prime} = \bm{J}\bm{W}\bm{\Sigma}\bm{W}^{T}\bm{J}^{T}, 
\label{eq:3dgs_proj}
\end{equation}
where $\bm{J}$ is the Jacobian matrix of the affine approximation of the projective transformation, $\bm{W}$ is the view transformation matrix. Due to the computation of transmittance $T_i$ in the Eq.~\ref{eq:3dgs_rendering}, the gaussians in scene are needed to be sorted according to the distance from the center point $\bm{\mu}$ of gaussian to camera focal point $\bm{f}$. After sorting the 3D Gaussians, we can render the image following by the Eq.~\ref{eq:3dgs_rendering}.

\subsection{Order-Independent Transparency Rendering}
\label{subsec:OIT}
Similar to 3DGS, OIT-based method represents the scene with $N$ 3D Gaussians $\mathcal{G}=\{\bm{\mu}_{i}, \bm{q}_i, \bm{s}_i, o_{i}, \bm{h}_i, w_i| i = 1, \ldots, N \}$, where $w$ indicates the learnable weight function, which is larger when gaussian near to camera. Different from the rendering equation in~\cite{kerbl3Dgaussians}, we use weighted OIT rendering equation, which is firstly defined in~\cite{mcguire2013weighted}:
\begin{equation}
\mathbf{C} = T\,\mathbf{c}_0 + (1-T)\,
\frac{\sum_{i=1}^{N} \mathbf{c}_i\,\alpha_i\,w_i}
     {\sum_{i=1}^{N} \alpha_i\,w_i},\quad T = \prod_{i=1}^{N} \left(1-\alpha_i\right)
\label{eq:oit_rendering}
\end{equation}
where the $\mathbf{c}_0$, $\mathbf{c}_i$ and $\alpha_i$ are background color, gaussian color and final opacity, respectively. We adopt the weight definition from~\cite{hou2024sort} and define $w_i$ as Eq.~\ref{eq:sf_weight}.

\section{Method}
\label{sec:method}
\begin{figure*}[t]
    \centering
    \includegraphics[width=1.0\linewidth]{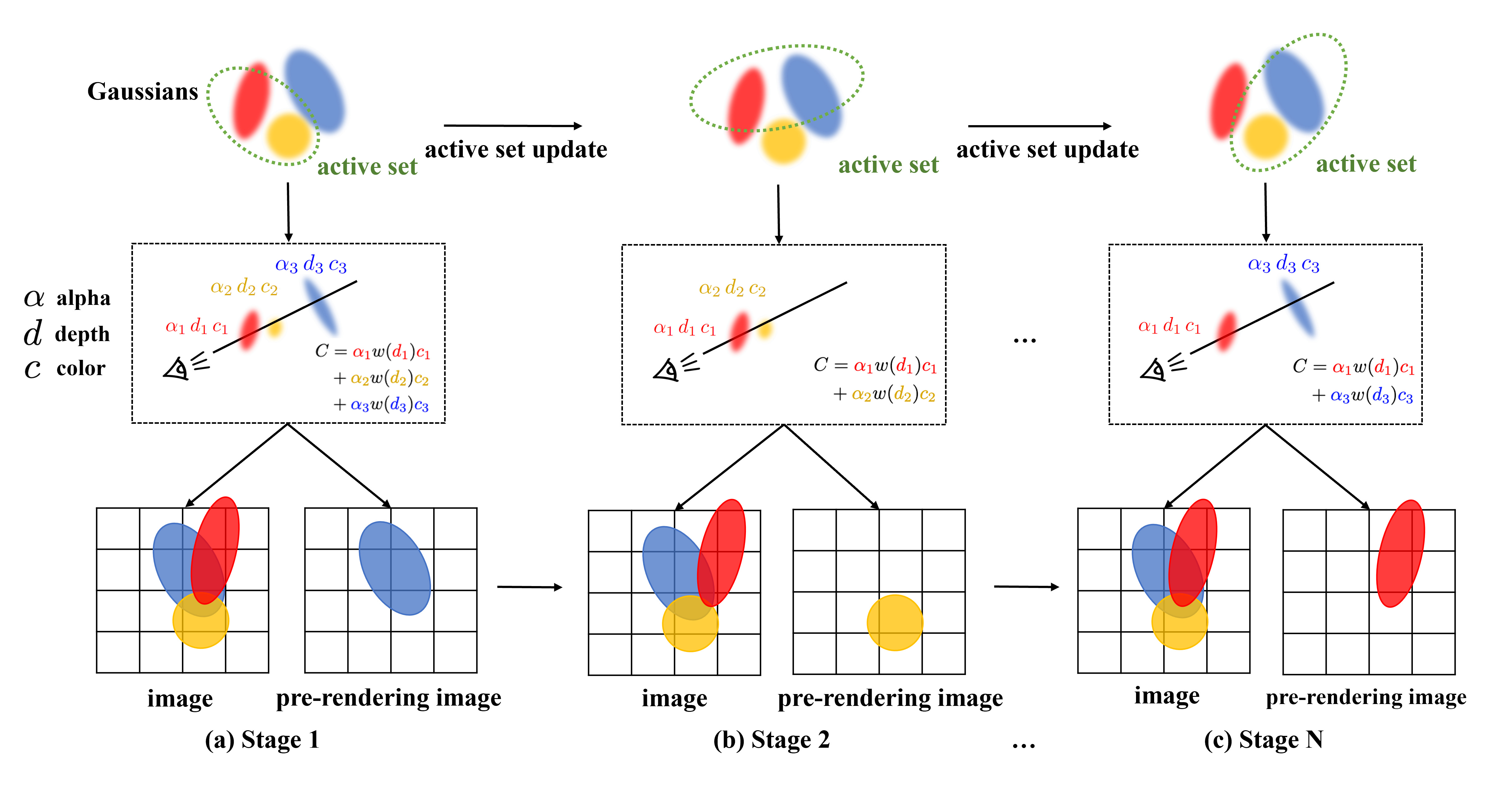}
    \caption{\textbf{Overview of SparseOIT.} Our SparseOIT framework takes multi-view images as input and follows the same optimization procedure as 3DGS in the early stage. After completing the first 15,000 iterations (including the densification phase), we switch to the proposed sparsity optimization with active set method.  We organize the optimization into stages separated by active set updates.  Before the each stage, we update the active set and use the pre-rendered result from the previous step (corresponding to the rendering of the inactive set). We then render images using the current active set, while simultaneously updating a new pre-rendered result that will serve as the input for the next stage.}
    \label{fig:overview}
\end{figure*}

We propose multiple strategies to improve OIT-based 3DGS reconstruction methods. In Sec.~\ref{subsec:act:set}, we first elaborate the acceleration framework via the active set method. We then introduce how we integrate the GPU acceleration techniques to significantly accelerate OIT-based methods in Sec.~\ref{subsec:integrate}. An overview of the active set acceleration pipeline is shown in Fig.~\ref{fig:overview}.

\subsection{Acceleration via Active Set Method}
\label{subsec:act:set}
Active set method works by iteratively finding \emph{active variables} that contributes the most to the optimization procedure. Instead of working on the full set of variables, active set method only optimize the active set of variables to reduce computational cost. There are three key factors that lead to the success of active set methods: 1) definition of activeness; 2) efficient update of active variables; and 3) efficient periodical update of the active set. For example, to make meaningful acceleration, the update of active variables and active set has to be done with a sublinear complexity, i.e., it should cost only a fraction of the computation time for computing a full-variable update. In the following, we elaborate how we design an active set method for OIT-based 3DGS reconstruction, focusing on the three key factors.

\noindent\textbf{Definition of Activeness.} Gaussian Splats are multi-dimensional variables consist of multiple attributes such as position, opacity and color. As these attributes are strongly correlated, we treat variables of the same gaussian as inseparable groups and only consider gaussian-level activeness. To determine the activeness of a gaussian $i$, we consider all its attributes $\{\bm{\mu}_i, \bm{q}_i, \bm{s}_i, o_i, \bm{h}_i, {w}_i\}$ and define the gaussian activeness w.r.t. the gradients of each variable:
\begin{equation}
    \bm{1}_{\textup{active}}(i) = \forall v \in \{\bm{\mu}_i, \bm{q}_i, \bm{s}_i, o_i, \bm{h}_i, w_i\},~(\|\nabla_v\| > \epsilon_v)
    \label{eq:activeness}
\end{equation}
In other words, the gaussian is active is at least one attribute variable is active, i.e., has a non-zero gradient. The $\epsilon_v$ for all attribute variables is empirically set to achieve proper balance between efficiency and performance.

\noindent\textbf{Updating Active Gaussians.} Given an active set $\A$ of gaussians and its complement set, the non-active set $\bar{\A}$, we adopt pre-rendering techniques to separately compute the rendering results for non-active gaussians and active gaussians. Since the OIT-based rendering equation does not couple gaussians, we can maintain a pre-rendered image for each training view. Between two consecutive updates of the active set, the pre-rendered image can be use to save computation on the non-active gaussians by fitting the rendering results of the active gaussians to the difference between the input image and the pre-rendered image. 

\noindent\textbf{Updating Active Set via Subsampling.} The active set $\A$ can be naively updated by re-evaluating the gradients of all Gaussians. However, this results in a computational complexity comparable to a full-variable update, thereby negating any potential acceleration. To reduce the cost of active set updates, we approximate the Gaussian gradients via subsampling. Since activeness is determined by comparing gradients with the non-zero vector, the decision process inherently tolerates a certain level of approximation error, which can be exploited for efficiency. To this end, during active set updates, we subsample a subset of viewpoints by applying farthest point sampling over the training views with a random initialization. Experimental results suggest that even with this subsampling strategy, the estimation of Gaussian activeness remains reliable. 

\noindent\textbf{Implementation Details.} After active set is updated, we also need to maintain the correctness of the pre-rendered images since the non-active set $\bar{\A}$ has changed. However, directly updating all pre-rendered images is prohibitive since the update is too sparse on the pre-rendered images, leading to poor GPU parallelism. Instead, we propose to delay the update of the pre-rendered image after each active set update until the pre-rendered image is used for training. We feed both the pre-rendered image and the training image into GPU and marks each gaussian by two categories of labels, indicating whether this gaussian is active or not, and process these two cases separately. Also, since the active set is almost full in the initial stage of the optimization, we choose to activate the active set method after $K=15,000$ iterations. The details of our active set algorithm are provided in Alg.~\ref{alg:oit_active}

\begin{algorithm}[b]
\caption{Active Set Method}
\label{alg:oit_active}
\begin{algorithmic}[1]  
\Require Training and Pre-rendered images $I, I^{\pre} \in R^{W \times H}$. Initial active set $\A$.
\Ensure  Gaussian parameters $\mathcal{G} = \{\bm{\mu}, \bm{q}, \bm{s}, o, \bm{h}, \bm{w}\}$
\State $\mathcal{G} \gets \text{GetAttributes()}$ 
\While{not converged}
  \State $I_i, I^{\pre}_i \gets \text{SampleTrainingView()}$ 
  \State $I^{\out}_i \gets \text{Rasterize}(\mathcal{G}_{\A}, I^{\pre}_i)$ 
  
  \State $L \gets \text{Loss}(I_i,\hat I^{\out}_i)$ 
  \State $\mathcal{G}_A \gets \text{ActiveGaussianUpdate}(\nabla L)$ 
 \If{$\text{IsActiveUpdateIteration}$}
        \For{each subsampled view $j$}
            \State $I^{\out}_j \gets \text{Rasterize}(\mathcal{G}, I^{\pre}_j)$ 
            \State $L_j \gets \text{Loss}(I^{\out}_j,I_j)$
        \EndFor
        \State $\A \gets \text{ActiveSetUpdate}(\{L_j\})$
  \EndIf
\EndWhile
\end{algorithmic}
\end{algorithm}

\subsection{Integrating GPU Acceleration Techniques}
\label{subsec:integrate}
As mentioned in~\cite{taminggs}, in the original 3DGS~\cite{kerbl3Dgaussians} backward propagation, gradients flow from pixels to Gaussians: each pixel back-propagates to every splat it touches, and the contributions are accumulated at the splat level via atomic operations. However, this design induces heavy contention as many threads target the same memory locations, causing GPU threads to stall on atomic writes for a large fraction of running time.

Similar in~\cite{taminggs}, we parallelize the backward propagation at the splat (rather than tile) level to mitigate contention from atomic gradient writes at the splat accumulators. However, because the original 3DGS~\cite{kerbl3Dgaussians} pipeline requires depth-sorted Gaussian rendering, it checkpoints intermediate per-pixel states every 32 splats during the forward pass and enforces a strict depth order traversal during backward propagation, thereby constraining the data access pattern. Although Taming-3DGS~\cite{taminggs} employs CUDA warp-shuffle operations to expedite state exchange, these ordering constraints induce up to 31 redundant computations per tile, due to the fixed warp size of 32 threads.

\begin{figure}[t]
    \centering
    \includegraphics[width=1.0\linewidth]{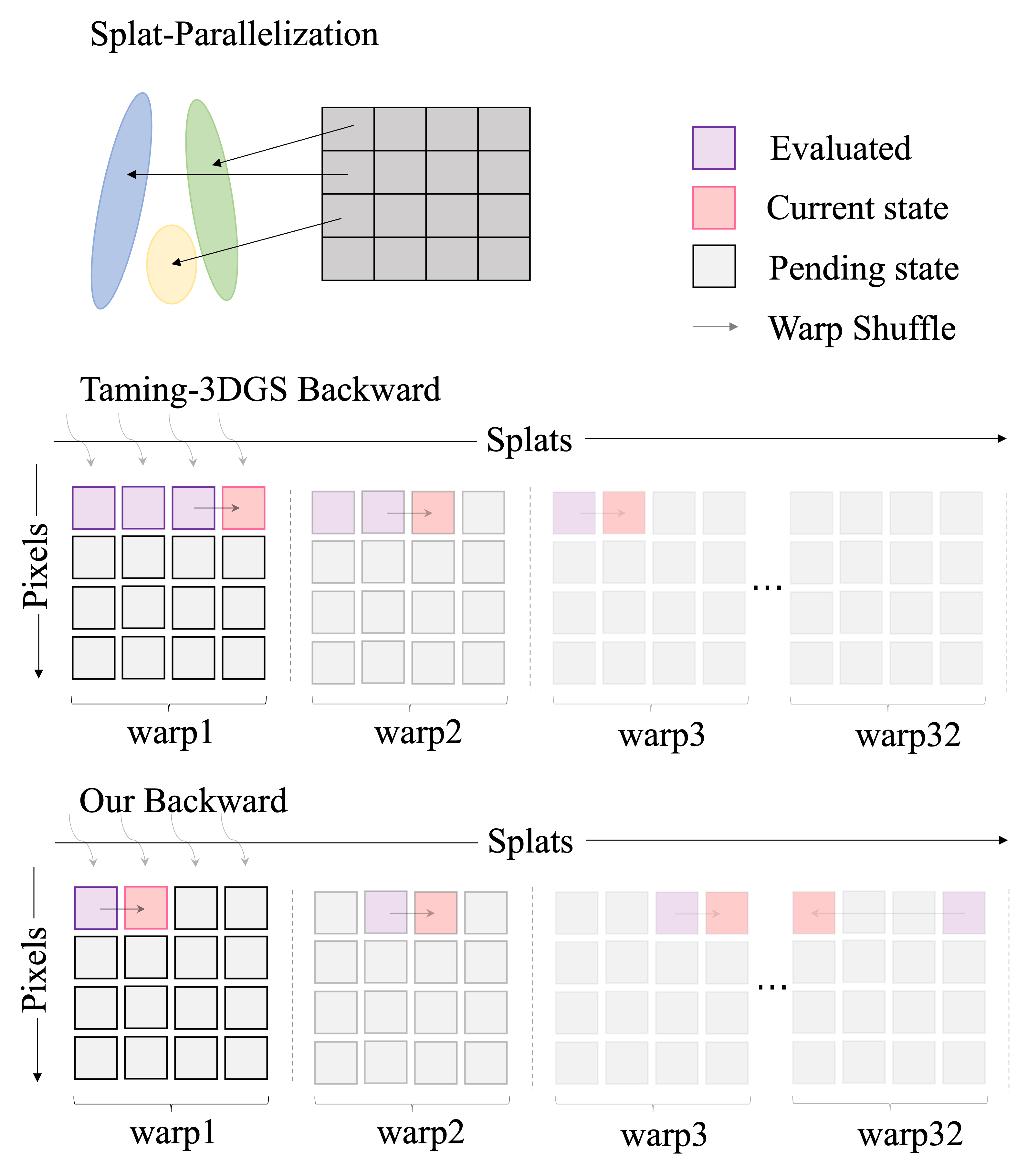}
    \caption{Gradient backpropagation. (Top) We utilize per-splat parallelization for backward propagation, which similar as Taming-3DGS~\cite{taminggs}. (Bottom) Taming-3DGS store the pixel states for every 32nd splat in the sorted list in forward propagation, while we need not do anything. For the backward, lane 0 of Taming-3DGS fetches each pixel’s parameters and broadcasts them to the entire warp. By contrast, our design allows every lane to fetch parameters for 32 pixels concurrently and exchange them across lanes, thereby improving efficiency.}
    \label{fig:recursive_backward_cuda}
\end{figure}

Benefiting from order-independent transparency (OIT), which imposes no constraint on the processing order of 3D Gaussians during rendering, we introduce a recursive per-splat parallelization for the backward propagation, as illustrated in Fig.~\ref{fig:recursive_backward_cuda}. Concretely, each tile (256 pixels) is partitioned into 8 groups of 32 pixels, matching the warp size. At the warp level, every lane fetches the value from its corresponding pixel; crucially, each group is fetched only once. We then circulate these per-pixel states within the warp via a loop, supplying each splat with 32 pixel states in turn and accumulating its gradients. Repeating this procedure eight times completes the tile-internal gradient computation and the associated atomic updates. This design enables simultaneous multi-lane fetching to reduce stalls while provably avoiding any redundant arithmetic. Additionally, we adopt culling strategy in~\cite{HansonSpeedy} to reduce the workload of both the forward and backward propagation, thereby avoiding the processing of redundant splats.

\section{Experiments}

\begin{figure*}[!t]
    \centering
    \includegraphics[width=\linewidth]{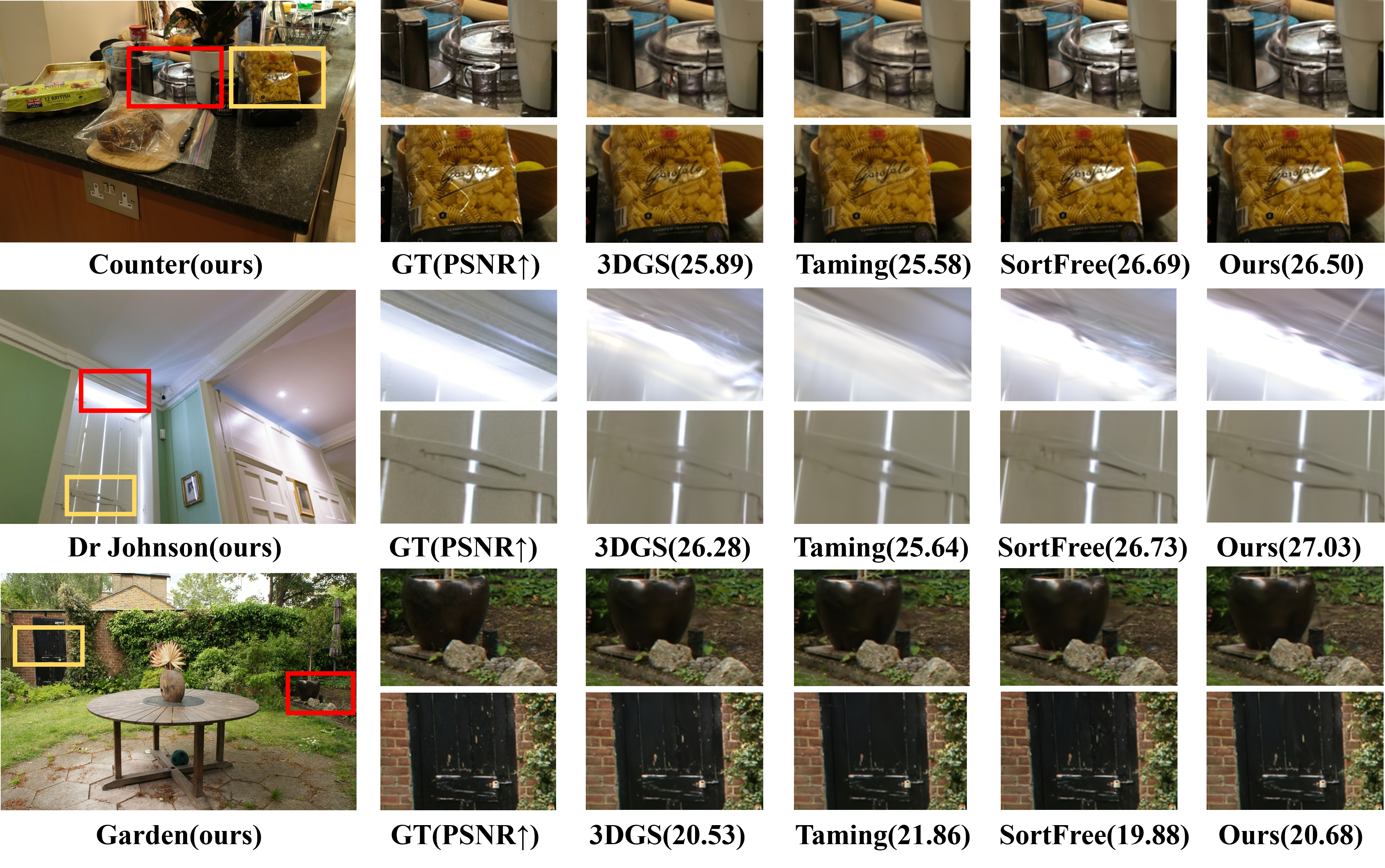}
    \caption{Comparison with baselines. We compare the SparseOIT-B (which incorporates the active set acceleration) against 3DGS~\cite{kerbl3Dgaussians}, Taming-3DGS~\cite{taminggs} (named “Taming” in the figure), and SortFree~\cite{hou2024sort}. For evaluation, we select indoor and outdoor scenes from the Mip-NeRF 360 dataset, as well as indoor scenes from the DeepBlending dataset. The PSNR values shown in this figure correspond to a single view.}
    \label{fig:experiment_result}
\end{figure*}

\label{sec:experiments}
We presents an experimental evaluation of proposed approach, including comparable evaluation by novel view synthesis, training efficiency, and the ablation study.
\subsection{Experimental Setup}

\noindent\textbf{Dataset.} 
We conduct experiments on three real-world datasets: Mip-NeRF 360~\cite{barron2022mipnerf360}, Deep-Blending~\cite{hedman2018deep}, and Tanks \& Temples~\cite{knapitsch2017tanks} for training and evaluation, which similar in~\cite{kerbl3Dgaussians}. Note that for the Mip-NeRF 360 scenes, we use images at $1/4$ of the original resolution used in 3DGS to prevent GPU memory overflow.

\noindent\textbf{Baselines.} 
We consider three baselines for comparsion. We quantitatively evaluate our method with 3DGS~\cite{kerbl3Dgaussians}, Taming-3DGS~\cite{taminggs}, and SortFree~\cite{hou2024sort}. SortFree is an OIT method, which do not need sorting to achieve 3D Gaussian rendering (Notably, we use a third-party implementation of SortFree, which has been acknowledged by the original authors on the first author’s personal homepage).

\noindent\textbf{Metrics.} We adopt the evaluation metrics used in~\cite{kerbl3Dgaussians}, including LPIPS~\cite{zhang2018unreasonable}, SSIM~\cite{wang2004image}, and PSNR (peak signal-to-noise ratio) to assess rendering quality. Additionally, we report the total training time in seconds to assess the training efficiency, sames in ~\cite{chen2025dashgaussian}. Furthermore, We also report the number of 3D Gaussians (expressed in thousands), which makes the comparison with the baselines more equitable, as the number of 3D Gaussian primitives has a strong impact on training efficiency.

\noindent\textbf{Computational Resources.} All experiments are executed on a single NVIDIA 4090 GPU with 24 GB GPU memory.

\noindent\textbf{Implementation Details.} We use Adam~\cite{adam2} for parameter optimization. The loss function is the same as in the original 3DGS formulation. The learning rates are set to $0.01$ for $o$, $0.1$ for $\sigma$, and $0.005$ for the $v$, with all other settings following the original 3DGS. Similarly with color $h$, $v$ is represented using single-channel third-order spherical harmonics. The loss weights $\lambda_{ssim}$ are also kept consistent with those of 3DGS. In addition, our densification strategy is kept consistent with that of 3DGS. To prevent GPU memory overflow, we set an empirical scene-specific random sampling probability during densification to limit the number of 3D Gaussians. Since the densification in 3DGS is terminated at 15,000 iterations, we start applying our active-set acceleration strategy from iteration 15,000. During the active set update stage, in order to avoid excessive computational overhead, we perform the update on a subset of 30 sampled views. To better preserve reconstruction quality, we select these views using a farthest point sampling strategy. In addition, our method exhibits noticeable variance across different runs. For some scenes, such as \textit{Bicycle}, \textit{Room}, and others, this can lead to a PSNR difference of up to 0.6 dB. Therefore, the values of the proposed method reported in the tables are averaged to provide a more reliable and representative evaluation.

\begin{table*}[t]
\centering
\small
\renewcommand{\arraystretch}{1}
\caption{Comparison with baselines on Tanks \& Temples, DeepBlending, and Mip-NeRF 360 datasets. SparseOIT-A denotes the variant without the active set acceleration, SparseOIT-B denotes the accelerate variant with active set method, and SparseOIT-C denotes the accelerated variant that adopts the densification strategy from Taming-3DGS~\cite{taminggs}(named “Taming” in the table). $N(k)$ denotes the number of Gaussians divided by 1000. Best and second-best results are highlighted in \bestbox\ best, \secondbox\ second best.}
\label{tab:tt_db_mip}

\begin{adjustbox}{scale=0.80}
\begin{tabular}{l ccccc ccccc ccccc}
\toprule
\multicolumn{1}{l}{\textbf{Method}}
& \multicolumn{5}{c}{\textbf{Tanks \& Temples}}
& \multicolumn{5}{c}{\textbf{DeepBlending}} 
& \multicolumn{5}{c}{\textbf{Mip-NeRF 360}} \\
\cmidrule(lr){2-6} \cmidrule(lr){7-11} \cmidrule(lr){12-16}
& PSNR$\uparrow$ & SSIM$\uparrow$ & LPIPS$\downarrow$ & Time$\downarrow$ & $N(k)$
& PSNR$\uparrow$ & SSIM$\uparrow$ & LPIPS$\downarrow$ & Time$\downarrow$ & $N(k)$
& PSNR$\uparrow$ & SSIM$\uparrow$ & LPIPS$\downarrow$ & Time$\downarrow$ & $N(k)$ \\
\midrule

3DGS~\cite{kerbl3Dgaussians}
& \best{23.78} & \best{0.8494} & \best{0.1704} & 705 & 1569
& 29.70 & 0.9027 & \second{0.2409} & 1213 & 2459
& \best{27.68} & \best{0.8214} & \best{0.1771} & 909 & 2679 \\

Taming~\cite{taminggs}
& \second{23.70} & 0.8320 & 0.2122 & \best{153} & 319
& 29.70 & 0.8992 & 0.2734 & \best{156} & 294
& \second{27.44} & 0.8012 & 0.2193 & \best{183} & 665 \\

SortFree~\cite{hou2024sort}
& 22.97 & 0.8299 & 0.1814 & 2159 & 3765
& 29.76 & 0.9016 & \best{0.2399} & 2065 & 2843
& 27.33 & \second{0.8067} & \second{0.1792} & 2302 & 4314 \\

SparseOIT-A
& 23.63 & 0.8422 & \second{0.1784} & 559 & 2055
& 29.76 & \second{0.9030} & 0.2479 & 355 & 1249
& 27.19 & 0.8019 & 0.2023 & 531 & 2126 \\

SparseOIT-B
& 23.68 & \second{0.8429} & 0.1798 & 445 & 2052
& \second{29.80} & \best{0.9043} & 0.2486 & 309 & 1251
& 27.21 & 0.8027 & 0.2040 & 408 & 2121 \\

SparseOIT-C
& 23.39 & 0.8206 & 0.2255 & \second{160} & 319
& \best{29.87} & 0.9010 & 0.2692 & \second{159} & 295
& 26.98 & 0.7802 & 0.2394 & \second{191} & 686 \\

\bottomrule
\end{tabular}
\end{adjustbox}
\end{table*}

\subsection{Comparison with Baselines}
In this section, we compare our method with three baselines. Our comparison primarily focuses on two aspects: novel view synthesis quality and training efficiency.

To enable a fair comparison of both novel view synthesis quality and training efficiency, we configure three variants of our SparseOIT method, among which SparseOIT-A denotes the variant that only employs the CUDA acceleration techniques described in Sec.\ref{subsec:integrate}, SparseOIT-B denotes the variant that incorporates both the CUDA acceleration techniques in Sec.\ref{subsec:integrate} and the active set strategy, and SparseOIT-C denotes the variant that employs the CUDA acceleration techniques in Sec.\ref{subsec:integrate} together with the densification strategy from Taming-3DGS~\cite{taminggs}, without using the active set acceleration. 
To enable a fair and comprehensive comparison with Taming-3DGS~\cite{taminggs}, we adopt the same densification strategy as in their approach and also employ their slow-update scheme for spherical harmonics. Moreover, we observe that under this strategy of Taming-3DGS~\cite{taminggs} introduce a different learning rate for the higher-order spherical harmonics residual terms. Moreover, since our method incorporates a spherical-harmonics-based weighting term, the slow color updating strategy of Taming-3DGS~\cite{taminggs} can be naturally extended and applied to this weight $v$ as well. 

In Fig.~\ref{fig:experiment_result}, we present the rendering comparison between SparseOIT-B (i.e., our proposed method) and other approaches. As shown in Fig.~\ref{fig:experiment_result}, OIT-based methods achieve better performance than 3DGS-based approaches on some scenes, especially for reflective objects. This is because, in the OIT formulation, the weights are optimized as independent variables, whereas in 3DGS the weights are strongly coupled with opacity and depth. In realistic indoor scenes with more complex lighting, particularly around reflective surfaces, this coupling makes it more difficult for 3DGS to faithfully model the appearance. 

As shown in Tab.~\ref{tab:tt_db_mip}, our SparseOIT-A and SparseOIT-B consistently outperform SortFree~\cite{hou2024sort} on both the Tanks \& Temples and DeepBlending datasets, while the performance gap on the Mip-NeRF 360 dataset remains relatively small. Moreover, their performance is comparable to that of 3DGS~\cite{kerbl3Dgaussians} and Taming-3DGS~\cite{taminggs}. In terms of training efficiency, our SparseOIT-B yields an approximately 4-6× speedup over SortFree~\cite{hou2024sort} and 2-4× speedup over 3DGS~\cite{kerbl3Dgaussians}. In addition, after adopting the same densification strategy as Taming-3DGS, our SparseOIT-C achieves a significantly higher training speed than both 3DGS and SortFree, and is comparable to Taming-3DGS~\cite{taminggs}. The remaining gap with Taming-3DGS mainly arises from differences in scene representation, the truncation used in the rendering process, and the fact that our model involves optimizing a slightly larger set of parameters than standard 3DGS/Taming-3DGS in same number of 3D Gaussian primitives, primarily due to the additional weighting coefficients. By optimizing the backward propagation pipeline on the CUDA side, we further narrow this gap. 

Furthermore, as reported in Tab.~\ref{tab:tt_db_mip}, SparseOIT-B consistently achieves higher training efficiency than SparseOIT-A, and the advantage becomes more pronounced as the number of 3D Gaussian primitives increases. This gain is enabled by the decoupling property of 3D Gaussian primitives under the OIT formulation, based on which we further introduce a pre-rendering strategy. In addition, our method supports more aggressive threshold settings, which further improve training efficiency.

\begin{table}[t]
\centering
\caption{The results of ablation study on backward propagation with recursive per-splat parallelization (Ours) with per-pixel backward propagation (3DGS) and per-splat backward propagation Taming-3DGS}
\label{tab:ablation}
\begin{adjustbox}{scale=0.8}
\begin{tabular}{l c c c c c}
\toprule
\textbf{Backward} 
& PSNR ↑ & SSIM ↑ & LPIPS ↓ & Time ↓ & N(k) \\
\midrule

Per-pixel~\cite{kerbl3Dgaussians}
& \textbf{30.04} & \textbf{0.9024} & \textbf{0.2474} & 981 & 1383 \\
Per-splat~\cite{taminggs}
& 29.93 & \textbf{0.9024} & 0.2477 & 415 & 1377 \\
Ours 
& 29.93 & 0.9020 & 0.2483 & \textbf{396} & 1380 \\

\bottomrule
\end{tabular}
\end{adjustbox}
\end{table}

\subsection{Ablation Study}
In this section, we conduct two ablation studies: (1) ablations on the backward propagation with recursive per-splat parallelization for Sec.~\ref{subsec:integrate}, and (2) ablations on the active set acceleration for Sec.~\ref{subsec:act:set}. In Tab.~\ref{tab:ablation}, we reports an ablation study on our backward-propagation optimizations. We conduct ablation experiments on the playroom scene from DeepBlending, and compare three variants: the 3DGS backward implementation, the Taming-3DGS backward implementation, and our own optimized backward implementation. As can be seen from the Tab.~\ref{tab:ablation}, our backward propagation with recursive
Per-Splat parallelization achieves roughly a 2× speedup over the original 3DGS, and yields an improvement of nearly 20 seconds compared to Taming-3DGS.

The impact of enabling the active set acceleration is summarized in Tab.~\ref{tab:tt_db_mip}. Specifically, SparseOIT-A denotes the variant without the active set strategy, whereas SparseOIT-B enables active set acceleration. From the results, we observe that introducing the active set strategy leads to clear improvements in training speed, while causing only negligible degradation in reconstruction quality. Moreover, the resulting speedup becomes increasingly pronounced as the number of 3D Gaussians grows, demonstrating the effectiveness of the proposed active set design in large-scale scenes. 

\section{Conclusion and Limitations}
\noindent\textbf{Conclusion.} In this work, we present an active-set-based sparse optimization framework for OIT-based 3D Gaussian Splatting, which effectively exploits the sparsity of Gaussian activeness during training and leads to substantial reductions in training time. We further integrate a set of practical GPU acceleration techniques into the algorithmic design, enabling our method to achieve performance comparable to state-of-the-art volumetric rendering approaches while significantly improving the efficiency of existing OIT-based methods. Extensive experiments and ablation studies validate the effectiveness of our design in terms of both rendering quality and training efficiency. 

\noindent\textbf{Limitations.} 
Pruning in OIT-based 3D Gaussian Splatting methods remains relatively under-explored, which can hinder the optimization of Gaussians and thereby affect both the final number of Gaussians and the overall training efficiency. Moreover, the design of the weight term 
$w$ introduces additional computational overhead and increased storage cost, and also impacts the quality of novel view synthesis, which partly accounts for the suboptimal rendering performance observed on certain scenes. These limitations highlight the importance of further investigating more effective pruning strategies, as well as alternative parameterizations of the weight $w$, in future work.
{
    \small
    \bibliographystyle{ieeenat_fullname}
    \bibliography{main}
}

\appendix
\clearpage
\setcounter{page}{1}
\maketitlesupplementary

\section{Details of Rasterizer with Pre-rendering}
\label{sec:Rasterizer}
In this section, we present the details of the Rasterizer with pre-rendering as Alg.~\ref{alg:rasterize}. The "BAN" means blend and normalize function. The "BAU" means blend and update function. It is worth noting that the sorting step only orders entries by tile ID and the blending process is independent of the depth order of the Gaussians.

\begin{algorithm}[h]
\caption{GPU software rasterization of 3D Gaussians}
\label{alg:rasterize}
\begin{algorithmic}[1]
\Require $w,h$; Gaussian means $M$ and covariances $S$; colors $C$ and opacities $A$; view $V$
\Ensure Rendered image $I$, pre-rendered image $I^{\pre}$
\State $\text{CullGaussian}(p, V)$  
\State $M',S' \gets \text{ScreenspaceGaussians}(M,S,V)$
\State $T \gets \text{CreateTiles}(w,h)$
\State $L,K \gets \text{DuplicateWithKeys}(M',T)$  
\State $ \text{SortByKeys}(K,L)$                    \Comment{Sort Only in Tile}
\State $R \gets \text{IdentifyTileRanges}(T,K)$
\State $I \gets 0$                                
\ForAll{tiles $t$ in $T$}
    \ForAll{pixels $i$ in $t$}
        \State $r \gets \text{GetTileRange}(R,t)$
        \State $I[i] \gets \text{BAN}(i,L,r,K,M',S',C,A,I^{\pre}[i])$
        \State $I^{\pre}[i] \gets \text{BAU}(i,L,r,K,M',S',C,A,I^{\pre}[i])$
    \EndFor
\EndFor
\State \Return $I$
\end{algorithmic}
\end{algorithm}

\section{Details of Backward Propagation}
\label{sec:backward}
In this section, we present the derivation of the backward propagation of SparseOIT. We rewrite the Eq.~\ref{eq:oit_rendering} as:
\begin{equation}
\begin{aligned}
\mathbf{C} &= T \mathbf{c}_0 + \left( 1-T \right)F,\\
F &= \frac{P}{Q},\quad P = \sum_{i=1}^{N} \mathbf{c}_i \alpha_i w_i,\quad Q = \sum_{i=1}^{N} \alpha_i w_i,
\label{eq:oit_rendering_md}
\end{aligned}
\end{equation}
Therefore, we can get the gradient of $\frac{\partial \mathbf{C}}{\partial \mathbf{c}_i}$, $\frac{\partial \mathbf{C}}{\partial \alpha_i}$, and $\frac{\partial \mathbf{C}}{\partial w_i}$ as:
\begin{equation}
\begin{aligned}
\frac{\partial \mathbf{C}}{\partial \mathbf{c}_i}
&= (1-T)\,\frac{\alpha_i w_i}{Q},\\
\frac{\partial \mathbf{C}}{\partial \alpha_i}
&= \frac{T}{1-\alpha_i}\,(F-\mathbf{c}_0)
  + (1-T)\,\frac{w_i}{Q}\,(\mathbf{c}_i - F),\\
\frac{\partial \mathbf{C}}{\partial w_i}
&= (1-T)\,\frac{\alpha_i}{Q}\,(\mathbf{c}_i - F).
\label{eq:oit_rendering_detail}
\end{aligned}
\end{equation}
From the backward propagation equation, we observe that the variables of each Gaussian are mutually independent. Therefore, we can apply the backward propagation with recursive Per-Splat parallelization strategy described in Sec.~\ref{subsec:integrate}, or alternatively adopt the active set method to further accelerate the computation.

\section{Analysis of View Subsampling Number and Update Interval for Active Set} 
In this section, we analyze the effects of the number of subsampled views and the update interval on the final performance. We conduct the corresponding experiments on the truck scene from the Tanks \& Temples dataset based on the SparseOIT-B variant. Tab.~\ref{tab:subsample} and Tab.~\ref{tab:interval} report the experimental results for the number of subsampled views and the update interval, respectively, where I denotes the update interval and S denotes the number of subsampled views. As shown in Tab.~\ref{tab:subsample}, increasing the number of subsampled views consistently leads to higher computational cost, while the rendering quality remains relatively stable with only minor fluctuations. Therefore, we use 30 subsampled views in our experiments, as this setting achieves a favorable trade-off between efficiency and performance. As shown in Tab.~\ref{tab:interval}, the training time generally increases as the update interval becomes longer. This is because a longer update interval slows down the reduction of active 3D Gaussians, resulting in higher computational cost during training. Therefore, we choose 500 as the update interval in most cases. For a few scenes, such as \textit{train}, \textit{room}, and \textit{bonsai}, however, we set the update interval to 600, which exceeds twice the number of training views, to ensure the correct execution of the algorithm.

\section{Detailed Evaluation Metrics} 
\label{sec:additional}
In this section, we further introduce SparseOIT-D, a variant that integrates the CUDA acceleration techniques described in Sec.~\ref{subsec:integrate}, the active set strategy, and the densification strategy from Taming-3DGS~\cite{taminggs}. The quantitative results are reported in Tab.~\ref{tab:tt_db_mip_modify}. In addition, we report per-scene PSNR, SSIM, LPIPS, training time, and the number of 3D Gaussians. Tab.~\ref{tab:PSNR_Result}, Tab.~\ref{tab:SSIM_Result}, Tab.~\ref{tab:LPIPS_Result} present the PSNR, SSIM, and LPIPS metrics for each scene within Mip-NeRF 360, Tanks \& Temples, and Deep Blending datasets, respectively. Tab.~\ref{tab:Time_Result} presents the per-scene training time across the three datasets, and Tab.~\ref{tab:Points_Result} provides the 3D Gaussian counts. 

\begin{table}[h]
\centering
\caption{Impact of the number of subsampled views on performance in the truck scene.}
\begin{adjustbox}{max width=\linewidth}
\begin{tabular}{l c c c c c}
\toprule
\textbf{Parameters} & PSNR $\uparrow$ & SSIM $\uparrow$ & LPIPS $\downarrow$ & Time $\downarrow$ & N(k) \\
\midrule
I=500, S=10  & 25.48 & 0.8745 & 0.1521 & 464 & 2307 \\
I=500, S=20  & 25.48 & 0.8753 & 0.1514 & 459 & 2285 \\
I=500, S=30  & 25.53 & 0.8753 & 0.1515 & 456 & 2288 \\
I=500, S=40  & 25.56 & 0.8752 & 0.1516 & 460 & 2279 \\
I=500, S=50  & 25.52 & 0.8750 & 0.1518 & 462 & 2290 \\
I=500, S=60  & 25.53 & 0.8753 & 0.1518 & 464 & 2283 \\
I=500, S=70  & 25.55 & 0.8753 & 0.1512 & 477 & 2294 \\
I=500, S=80  & 25.51 & 0.8749 & 0.1524 & 471 & 2282 \\
I=500, S=90  & 25.48 & 0.8748 & 0.1524 & 470 & 2269 \\
I=500, S=100 & 25.55 & 0.8755 & 0.1519 & 472 & 2255 \\
\bottomrule
\end{tabular}
\end{adjustbox}
\label{tab:subsample}
\end{table}

\begin{table}[h]
\centering
\caption{Impact of the update intervals on performance in the truck scene.}
\begin{adjustbox}{max width=\linewidth}
\begin{tabular}{l c c c c c}
\toprule
\textbf{Parameters} & PSNR $\uparrow$ & SSIM $\uparrow$ & LPIPS $\downarrow$ & Time $\downarrow$ & N(k) \\
\midrule
I=500, S=30   & 25.53 & 0.8753 & 0.1515 & 456 & 2288 \\
I=600, S=30   & 25.57 & 0.8755 & 0.1516 & 461 & 2290 \\
I=700, S=30   & 25.51 & 0.8746 & 0.1516 & 455 & 2296 \\
I=800, S=30   & 25.52 & 0.8753 & 0.1523 & 459 & 2261 \\
I=900, S=30   & 25.52 & 0.8754 & 0.1524 & 459 & 2251 \\
I=1000, S=30  & 25.54 & 0.8752 & 0.1518 & 464 & 2275 \\
I=2000, S=30  & 25.52 & 0.8747 & 0.1530 & 480 & 2285 \\
I=3000, S=30  & 25.52 & 0.8750 & 0.1521 & 492 & 2257 \\
I=4000, S=30  & 25.51 & 0.8754 & 0.1517 & 503 & 2265 \\
I=5000, S=30  & 25.58 & 0.8755 & 0.1520 & 516 & 2274 \\
\bottomrule
\end{tabular}
\end{adjustbox}
\label{tab:interval}
\end{table}

\section{Further Analysis and Discussion} 
From our empirical observations, the outdoor scenes in the Mip-NeRF 360  dataset tend to overfit under the learning rates adopted in the main paper. To further investigate this issue, we additionally conduct experiments with the learning rate of parameter $v$ reduced to $0.0005$, while keeping all other settings unchanged. Following the experimental setting in Sec.~\ref{sec:additional}, we further provide both the average results on the Mip-NeRF 360 dataset in Tab.~\ref{tab:mip_nerf} and the detailed per-scene results for the outdoor scenes in Tab.~\ref{tab:PSNR_SSIM_LPIPS_Result} and Tab.~\ref{tab:Time_Count_Result}.

It can be observed that the rendering quality on the outdoor scenes of Mip-NeRF 360 improves substantially. This phenomenon may be due to the fact that the current weight formulation of 3D Gaussians is not sufficiently accurate, which also makes the optimization more difficult. Moreover, the slightly worse performance of the current OIT formulation compared with 3DGS may also stem from the relatively lower quality metrics on the outdoor scenes of the Mip-NeRF 360 dataset.

In addition, empirical results from our initial experiments suggest that the random sampling strategy applied during densification can improve novel view synthesis performance on the evaluation set across multiple scenes, with representative examples including \textit{train}, \textit{drjohnson}, and \textit{treehill}. This may be attributed to the fact that a larger number of 3D Gaussians makes the scene more prone to overfitting, thereby degrading rendering quality on unseen views.

\begin{table}[h]
\centering
\caption{Quantitative results on Mip-NeRF 360 under adjusted learning rate}
\label{tab:mip_nerf}
\resizebox{\linewidth}{!}{%
\begin{tabular}{cccccc}
\toprule
\textbf{Method} & PSNR$\uparrow$ & SSIM$\uparrow$ & LPIPS$\downarrow$ & Time$\downarrow$ & N(k) \\
\midrule
3DGS~\cite{kerbl3Dgaussians} 
& \best{27.68} & \best{0.8214} & \best{0.1771} & 909 & 2679 \\

Taming-3DGS~\cite{taminggs} 
& 27.44 & 0.8012 & 0.2193 & \second{183} & 665 \\

SortFree~\cite{hou2024sort} 
& 27.33 & \second{0.8067} & \second{0.1792} & 2302 & 4314 \\

SparseOIT-A 
& \second{27.38} & 0.8058 & 0.1964 & 564 & 2082 \\

SparseOIT-B 
& \second{27.38} & 0.8058 & 0.1973 & 460 & 2094 \\

SparseOIT-C 
& 27.11 & 0.7884 & 0.2268 & 201 & 686 \\

SparseOIT-D 
& 26.92 & 0.7805 & 0.2377 & \best{175} & 686 \\

\bottomrule
\end{tabular}
}
\end{table}

\begin{table*}[t]
\centering
\small
\renewcommand{\arraystretch}{1}
\caption{Comparison with baselines on Tanks \& Temples, DeepBlending, and Mip-NeRF datasets. $N(k)$ denotes the number of Gaussians divided by 1000. Best and second-best results are highlighted in \bestbox\ best, \secondbox\ second best.}
\label{tab:tt_db_mip_modify}

\begin{adjustbox}{scale=0.75}
\begin{tabular}{l ccccc ccccc ccccc}
\toprule
\multirow{2}{*}{\textbf{Method}}
& \multicolumn{5}{c}{\textbf{Tanks \& Temples}}
& \multicolumn{5}{c}{\textbf{DeepBlending}} 
& \multicolumn{5}{c}{\textbf{Mip-NeRF 360}} \\
\cmidrule(lr){2-6} \cmidrule(lr){7-11} \cmidrule(lr){12-16}
& PSNR$\uparrow$ & SSIM$\uparrow$ & LPIPS$\downarrow$ & Time$\downarrow$ & $N(k)$
& PSNR$\uparrow$ & SSIM$\uparrow$ & LPIPS$\downarrow$ & Time$\downarrow$ & $N(k)$ 
& PSNR$\uparrow$ & SSIM$\uparrow$ & LPIPS$\downarrow$ & Time$\downarrow$ & $N(k)$  \\
\midrule

3DGS~\cite{kerbl3Dgaussians}
& \best{23.78} & \best{0.8494} & \best{0.1704} & 705 & 1569
& 29.70 & 0.9027 & \second{0.2409} & 1213 & 2459
& \best{27.68} & \best{0.8214} & \best{0.1771} & 909 & 2679 \\

Taming-3DGS~\cite{taminggs}
& \second{23.70} & 0.8320 & 0.2122 & \best{153} & 319
& 29.70 & 0.8992 & 0.2734 & \second{156} & 294
& \second{27.44} & 0.8012 & 0.2193 & \second{183} & 665 \\

SortFree~\cite{hou2024sort}
& 22.97 & 0.8299 & 0.1814 & 2159 & 3765
& 29.76 & 0.9016 & \best{0.2399} & 2065 & 2843
& 27.33 & \second{0.8067} & \second{0.1792} & 2302 & 4314 \\

SparseOIT-A
& 23.63 & 0.8422 & \second{0.1784} & 559 & 2055
& 29.76 & \second{0.9030} & 0.2479 & 355 & 1249
& 27.19 & 0.8019 & 0.2023 & 531 & 2126 \\

SparseOIT-B
& 23.68 & \second{0.8429} & 0.1798 & 445 & 2052
& 29.80 & \best{0.9043} & 0.2486 & 309 & 1251
& 27.21 & 0.8027 & 0.2040 & 408 & 2121 \\

SparseOIT-C
& 23.39 & 0.8206 & 0.2255 & 160 & 319
& \best{29.87} & 0.9010 & 0.2692 & 159 & 295
& 26.98 & 0.7802 & 0.2394 & 191 & 686 \\

SparseOIT-D
& 23.34 & 0.8143 & 0.2374 & \second{159} & 319
& \second{29.84} & 0.8993 & 0.2783 & \best{150} & 295
& 26.84 & 0.7737 & 0.2500 & \best{170} & 686 \\

\bottomrule
\end{tabular}
\end{adjustbox}
\end{table*}

\begin{table*}[t]
\centering
\small
\renewcommand{\arraystretch}{1}
\caption{PSNR scores of our method on the Mip-NeRF 360 dataset, the Tanks \& Temples dataset, and the Deep Blending dataset.}
\label{tab:PSNR_Result}

\begin{adjustbox}{scale=0.80}
\begin{tabular}{l cccccccccc ccc ccc}
\toprule
\multirow{1}{*}{\textbf{Method}}
& \multicolumn{9}{c}{\textbf{Mip-NeRF 360}}
& \multicolumn{2}{c}{\textbf{Tanks \& Temples}} 
& \multicolumn{2}{c}{\textbf{Deep Blending}} \\
\cmidrule(lr){2-10} \cmidrule(lr){11-12} \cmidrule(lr){13-14}
& bicycle & flowers & garden & stump & treehill & room & counter & kitchen & bonsai 
& truck & train  
& drjohnson & playroom & \\
\midrule

3DGS~\cite{kerbl3Dgaussians}
& 25.24 & 21.50 & 27.37 & 26.64 & 22.54 & 32.23 & 29.37 & 31.80 & 32.47 & 25.44
& 22.12 & 29.41 & 30.00 \\

Taming-3DGS~\cite{taminggs}
& 24.85 & 21.01 & 27.25 & 25.99 & 22.89 & 32.14 & 28.91 & 31.65 & 32.26 & 25.17
& 22.24 & 29.43 & 29.98 \\

SortFree~\cite{hou2024sort}
& 23.89 & 20.58 & 27.07 & 24.75 & 21.09 & 32.90 & 30.30 & 31.90 & 33.45 & 24.58
& 21.36 & 29.51 & 30.00 \\

SparseOIT-A
& 23.92 & 20.42 & 26.95 & 24.85 & 22.32 & 32.11 & 29.88 & 31.31 & 32.93 & 25.45
& 21.82 & 29.59 & 29.93 \\

SparseOIT-B
& 23.86 & 20.49 & 27.11 & 25.02 & 22.34 & 32.13 & 29.83 & 31.33 & 32.78 & 25.53
& 21.82 & 29.64 & 29.95 \\

SparseOIT-C
& 23.53 & 20.11 & 26.95 & 24.64 & 22.40 & 31.87 & 29.36 & 31.21 & 32.74 & 24.95
& 21.85 & 29.56 & 30.18 \\

SparseOIT-D
& 23.66 & 20.02 & 26.53 & 24.74 & 22.53 & 31.55 & 29.31 & 30.72 & 32.48 & 24.96
& 21.73 & 29.54 & 30.13 \\

\bottomrule
\end{tabular}
\end{adjustbox}
\end{table*}

\begin{table*}[t]
\centering
\small
\renewcommand{\arraystretch}{1}
\caption{SSIM scores of our method on the Mip-NeRF360 dataset, the Tanks \& Temples dataset, and
the Deep Blending dataset.}
\label{tab:SSIM_Result}

\begin{adjustbox}{scale=0.80}
\begin{tabular}{l cccccccccc ccc ccc}
\toprule
\multirow{1}{*}{\textbf{Method}}

& \multicolumn{9}{c}{\textbf{Mip-NeRF 360}}
& \multicolumn{2}{c}{\textbf{Tanks \& Temples}} 
& \multicolumn{2}{c}{\textbf{Deep Blending}} \\
\cmidrule(lr){2-10} \cmidrule(lr){11-12} \cmidrule(lr){13-14}

& bicycle & flowers & garden & stump & treehill & room & counter & kitchen & bonsai 
& truck & train 
& drjohnson & playroom  \\
\midrule

3DGS~\cite{kerbl3Dgaussians}
& 0.7647 & 0.6029 & 0.8640 & 0.7714 & 0.6326 & 0.9415 & 0.9176 & 0.9444 & 0.9532 & 0.8810
& 0.8178 & 0.9026 & 0.9028 \\

Taming-3DGS~\cite{taminggs}
& 0.7169 & 0.5531 & 0.8532 & 0.7351 & 0.6227 & 0.9327 & 0.9084 & 0.9406 & 0.9485 & 0.8638
& 0.8002 & 0.8996 & 0.8987 \\

SortFree~\cite{hou2024sort}
& 0.7281 & 0.5875 & 0.8570 & 0.7054 & 0.6126 & 0.9448 & 0.9241 & 0.9444 & 0.9563 & 0.8665
& 0.7933 & 0.9012 & 0.9020 \\

SparseOIT-A
& 0.6942 & 0.5743 & 0.8492 & 0.7181 & 0.6219 & 0.9430 & 0.9205 & 0.9422 & 0.9534 & 0.8753
& 0.8091 & 0.9040 & 0.9020 \\

SparseOIT-B
& 0.6952 & 0.5751 & 0.8504 & 0.7233 & 0.6218 & 0.9429 & 0.9201 & 0.9423 & 0.9531 & 0.8753
& 0.8106 & 0.9050 & 0.9037 \\

SparseOIT-C
& 0.6406 & 0.5319 & 0.8379 & 0.6822 & 0.5970 & 0.9336 & 0.9117 & 0.9375 & 0.9495 & 0.8532
& 0.7880 & 0.9006 & 0.9015 \\

SparseOIT-D
& 0.6347 & 0.5189 & 0.8220 & 0.6778 & 0.5917 & 0.9299 & 0.9071 & 0.9344 & 0.9466 & 0.8527
& 0.7758 & 0.8985 & 0.9001 \\

\bottomrule
\end{tabular}
\end{adjustbox}
\end{table*}

\begin{table*}[t]
\centering
\small
\renewcommand{\arraystretch}{1}
\caption{LPIPS scores of our method on the Mip-NeRF360 dataset, the Tanks \& Temples dataset, and
the Deep Blending dataset.}
\label{tab:LPIPS_Result}

\begin{adjustbox}{scale=0.80}
\begin{tabular}{l cccccccccc ccc ccc}
\toprule
\multirow{1}{*}{\textbf{Method}}
& \multicolumn{9}{c}{\textbf{Mip-NeRF 360}}
& \multicolumn{2}{c}{\textbf{Tanks \& Temples}} 
& \multicolumn{2}{c}{\textbf{Deep Blending}} \\
\cmidrule(lr){2-10} \cmidrule(lr){11-12} \cmidrule(lr){13-14}

& bicycle & flowers & garden & stump & treehill & room & counter & kitchen & bonsai 
& truck & train 
& drjohnson & playroom  \\
\midrule

3DGS~\cite{kerbl3Dgaussians}
& 0.2105 & 0.3379 & 0.1078 & 0.2153 & 0.3287 & 0.1140 & 0.1143 & 0.0677 & 0.0974 & 0.1428
& 0.1980 & 0.2388 & 0.2429 \\

Taming-3DGS~\cite{taminggs}
& 0.2970 & 0.4088 & 0.1287 & 0.2924 & 0.3892 & 0.1366 & 0.1320 & 0.0759 & 0.1132 & 0.1854
& 0.2390 & 0.2692 & 0.2777 \\

SortFree~\cite{hou2024sort}
& 0.2352 & 0.3101 & 0.1142 & 0.2510 & 0.3304 & 0.1081 & 0.1046 & 0.0677 & 0.0916 & 0.1509
& 0.2118 & 0.2364 & 0.2433 \\

SparseOIT-A
& 0.3098 & 0.3658 & 0.1332 & 0.2556 & 0.3652 & 0.1116 & 0.1110 & 0.0704 & 0.0983 & 0.1502
& 0.2067 & 0.2474 & 0.2483 \\

SparseOIT-B
& 0.3135 & 0.3682 & 0.1351 & 0.2561 & 0.3697 & 0.1126 & 0.1124 & 0.0704 & 0.0978 & 0.1515
& 0.2081 & 0.2487 & 0.2485 \\

SparseOIT-C
& 0.3711 & 0.4218 & 0.1559 & 0.3413 & 0.4117 & 0.1351 & 0.1301 & 0.0793 & 0.1082 & 0.2005
& 0.2505 & 0.2633 & 0.2750 \\

SparseOIT-D
& 0.3829 & 0.4322 & 0.1836 & 0.3518 & 0.4245 & 0.1420 & 0.1364 & 0.0834 & 0.1132 & 0.2031
& 0.2718 & 0.2750 & 0.2817 \\
\bottomrule
\end{tabular}
\end{adjustbox}
\end{table*}

\begin{table*}[t]
\centering
\small
\renewcommand{\arraystretch}{1}
\caption{Training time of our method on the Mip-NeRF360 dataset, the Tanks \& Temples dataset, and the Deep Blending dataset.}
\label{tab:Time_Result}

\begin{adjustbox}{scale=0.80}
\begin{tabular}{l cccccccccc ccc ccc}
\toprule
\multirow{1}{*}{\textbf{Method}}
& \multicolumn{9}{c}{\textbf{Mip-NeRF 360}}
& \multicolumn{2}{c}{\textbf{Tanks \& Temples}} 
& \multicolumn{2}{c}{\textbf{Deep Blending}} \\
\cmidrule(lr){2-10} \cmidrule(lr){11-12} \cmidrule(lr){13-14}
& bicycle & flowers & garden & stump & treehill & room & counter & kitchen & bonsai 
& truck & train 
& drjohnson & playroom  \\
\midrule

3DGS~\cite{kerbl3Dgaussians}
3DGS~\cite{kerbl3Dgaussians}
& 1467 & 975 & 1396 & 1207 & 1082 & 495 & 494 & 621 & 443 & 776
& 634 & 1396 & 1031 \\

Taming-3DGS~\cite{taminggs}
& 206 & 178 & 324 & 156 & 199 & 129 & 144 & 162 & 145 & 144
& 161 & 165 & 147 \\

SortFree~\cite{hou2024sort}
& 3423 & 3494 & 3063 & 3251 & 3198 & 1037 & 1159 & 1129 & 965 & 2076
& 2242 & 2640 & 1491 \\

SparseOIT-A
& 523 & 500 & 817 & 933 & 497 & 333 & 373 & 430 & 376 & 612
& 505 & 314 & 396 \\

SparseOIT-B
& 417 & 411 & 603 & 607 & 409 & 262 & 297 & 363 & 303 & 456
& 434 & 288 & 329 \\

SparseOIT-C
& 199 & 170 & 352 & 163 & 201 & 140 & 158 & 176 & 165 & 154
& 166 & 171 & 146 \\

SparseOIT-D
& 167 & 164 & 241 & 158 & 176 & 134 & 158 & 175 & 154 & 157
& 162 & 159 & 141 \\
\bottomrule
\end{tabular}
\end{adjustbox}
\end{table*}

\begin{table*}[!t]
\vspace{-1.20in}
\centering
\small
\renewcommand{\arraystretch}{1}
\caption{Per-scene 3D Gaussian counts of our method on the Mip-NeRF 360, Tanks \& Temples, and Deep Blending datasets, where each value is reported as $N(k)$ , the number of Gaussians divided by 1000.}
\label{tab:Points_Result}

\begin{adjustbox}{scale=0.80}
\begin{tabular}{l cccccccccc ccc ccc}
\toprule
\multirow{1}{*}{\textbf{Method}}
& \multicolumn{9}{c}{\textbf{Mip-NeRF 360}}
& \multicolumn{2}{c}{\textbf{Tanks \& Temples}} 
& \multicolumn{2}{c}{\textbf{Deep Blending}} \\
\cmidrule(lr){2-10} \cmidrule(lr){11-12} \cmidrule(lr){13-14}
& bicycle & flowers & garden & stump & treehill & room & counter & kitchen & bonsai 
& truck & train 
& drjohnson & playroom  \\
\midrule

3DGS~\cite{kerbl3Dgaussians}
& 4876 & 2865 & 4069 & 4327 & 3269 & 1141 & 994 & 1509 & 1064 & 2054
& 1084 & 3076 & 1841 \\

Taming-3DGS~\cite{taminggs}
& 813 & 575 & 1903 & 480 & 785 & 225 & 311 & 482 & 413 & 272
& 365 & 404 & 185 \\

SortFree~\cite{hou2024sort}
& 6312 & 7296 & 5238 & 6498 & 6269 & 1676 & 2076 & 1872 & 1591 & 3694
& 3835 & 3650 & 2035 \\

SparseOIT-A
& 2293 & 2159 & 3227 & 4204 & 2124 & 1133 & 1310 & 1414 & 1268 & 2275
& 1835 & 1118 & 1380 \\

SparseOIT-B
& 2261 & 2136 & 3232 & 4191 & 2110 & 1140 & 1322 & 1426 & 1271 & 2288
& 1817 & 1118 & 1384 \\

SparseOIT-C
& 814 & 575 & 2081 & 483 & 785 & 225 & 312 & 483 & 413 & 272
& 366 & 404 & 185 \\

SparseOIT-D
& 814 & 575 & 2081 & 483 & 785 & 225 & 311 & 483 & 413 & 272
& 366 & 404 & 185 \\

\bottomrule
\end{tabular}
\end{adjustbox}
\end{table*}

\begin{table*}[t]
\vspace{-3.5in}
\centering
\small
\renewcommand{\arraystretch}{1}
\caption{PSNR, SSIM and LPIPS scores of our method in the outdoor scenes of Mip-NeRF 360 dataset under adjusted learning rate}
\label{tab:PSNR_SSIM_LPIPS_Result}

\begin{adjustbox}{scale=0.75}
\begin{tabular}{l cccccccccc ccc ccc}
\toprule
\multirow{1}{*}{\textbf{Method}}
& \multicolumn{5}{c}{\textbf{PSNR}}
& \multicolumn{5}{c}{\textbf{SSIM}} 
& \multicolumn{5}{c}{\textbf{LPIPS}} \\
\cmidrule(lr){2-6} \cmidrule(lr){7-11} \cmidrule(lr){12-16}
& bicycle & flowers & garden & stump & treehill 
& bicycle & flowers & garden & stump & treehill
& bicycle & flowers & garden & stump & treehill\\
\midrule

3DGS~\cite{kerbl3Dgaussians}
& 25.24 & 21.50 & 27.37
& 26.64 & 22.54 & 0.7647
& 0.6029 & 0.8640 & 0.7714
& 0.6326 & 0.2105 & 0.3379
& 0.1078 & 0.2153 & 0.3287 \\

Taming-3DGS~\cite{taminggs}
& 24.85 & 21.01 & 27.25
& 25.99 & 22.89 & 0.7169
& 0.5531 & 0.8532 & 0.7351
& 0.6227 & 0.2970 & 0.4088
& 0.1287 & 0.2924 & 0.3892 \\

SortFree~\cite{hou2024sort}
& 23.89 & 20.58 & 27.07
& 24.75 & 21.09 & 0.7281
& 0.5875 & 0.8570 & 0.7054
& 0.6126 & 0.2352 & 0.3101
& 0.1142 & 0.2510 & 0.3304 \\

SparseOIT-A
& 24.25 & 20.72 & 27.13
& 25.42 & 22.63 & 0.7239
& 0.5635 & 0.8573 & 0.7247
& 0.6239 & 0.2705 & 0.3734
& 0.1168 & 0.2521 & 0.3638 \\

SparseOIT-B
& 24.35 & 20.78 & 27.13
& 25.44 & 22.65 & 0.7244
& 0.5629 & 0.8570 & 0.7268
& 0.6229 & 0.2720 & 0.3740
& 0.1182 & 0.2518 & 0.3665 \\

SparseOIT-C
& 23.85 & 20.19 & 26.95
& 25.11 & 22.72 & 0.6870
& 0.5199 & 0.8490 & 0.6992
& 0.6080 & 0.3218 & 0.4233
& 0.1304 & 0.3167 & 0.3960 \\

SparseOIT-D
& 23.72 & 20.10 & 26.75
& 25.11 & 22.55 & 0.6748
& 0.5028 & 0.8408 & 0.6914
& 0.5968 & 0.3389 & 0.4376
& 0.1437 & 0.3301 & 0.4142 \\

\bottomrule
\end{tabular}
\end{adjustbox}
\end{table*}

\begin{table*}[t]
\vspace{-3.5in}
\centering
\small
\renewcommand{\arraystretch}{1}
\caption{Training time and per-scene 3D Gaussian counts of our method in the outdoor scenes of Mip-NeRF 360 dataset under adjusted learning rate}
\label{tab:Time_Count_Result}

\begin{adjustbox}{scale=0.80}
\begin{tabular}{l cccccccccc ccc ccc}
\toprule
\multirow{1}{*}{\textbf{Method}}
& \multicolumn{5}{c}{\textbf{Time}}
& \multicolumn{5}{c}{\textbf{N(k)}} \\
\cmidrule(lr){2-6} \cmidrule(lr){7-11} \cmidrule(lr){12-16}
& bicycle & flowers & garden & stump & treehill 
& bicycle & flowers & garden & stump & treehill \\
\midrule

3DGS~\cite{kerbl3Dgaussians} 
& 1467 & 975
& 1396 & 1207
& 1082 & 4876
& 2865 & 4069
& 4327 & 3269 \\

Taming-3DGS~\cite{taminggs} 
& 206 & 178
& 324 & 156
& 199 & 813
& 575 & 1903
& 480 & 785 \\

SortFree~\cite{hou2024sort} 
& 3423 & 3494
& 3063 & 3251
& 3198 & 6312
& 7296 & 5238
& 6498 & 6269 \\

SparseOIT-A 
& 574 & 585
& 908 & 980
& 513 & 2345
& 2198 & 3206
& 3821 & 2041 \\

SparseOIT-B 
& 493 & 491
& 731 & 735
& 461 & 2349
& 2177 & 3212
& 3904 & 2043 \\

SparseOIT-C 
& 220 & 187
& 383 & 171
& 213 & 814
& 575 & 2081
& 484 & 785 \\

SparseOIT-D 
& 177 & 169
& 264 & 156
& 188 & 814
& 575 & 2081
& 484 & 785 \\

\bottomrule
\end{tabular}
\end{adjustbox}
\end{table*}


\end{document}